\documentclass[conference]{IEEEtran}
\IEEEoverridecommandlockouts
\usepackage{subfig}
\usepackage{cite}
\usepackage{amsmath,amssymb,amsfonts}
\usepackage{algorithmic}
\usepackage{graphicx}
\usepackage{textcomp}
\usepackage{xcolor}
\usepackage{algorithm,algorithmic}
\usepackage{amsthm}
\usepackage{bm}
\usepackage{comment}

\newcommand{\bl}{\textcolor{black}}

\def\BibTeX{{\rm B\kern-.05em{\sc i\kern-.025em b}\kern-.08em
    T\kern-.1667em\lower.7ex\hbox{E}\kern-.125emX}}
\begin{document}

\title{Digital Twin-Based User-Centric Edge Continual Learning in Integrated Sensing and Communication}

\author{Shisheng Hu$^{\ast}$, Jie Gao$^{\dag}$, Xinyu Huang$^{\ast}$, Mushu Li$^{\ddagger}$, Kaige Qu$^{\ast}$, Conghao Zhou$^{\ast}$, and Xuemin (Sherman) Shen$^{\ast}$\\
{\normalsize $^{\ast}$ Department of Electrical and Computer Engineering, University of Waterloo, Canada} \\
{\normalsize$^{\dag}$School of Information Technology, Carleton University, Canada}\\
{\normalsize$^{\ddagger}$ Department of Electrical, Computer, and Biomedical Engineering, Toronto Metropolitan University, Canada}\\
{\normalsize Email: \{s97hu, x357huan, k2qu, c89zhou, sshen\}@uwaterloo.ca, jie.gao6@carleton.ca, mushu1.li@ryerson.ca}}
\maketitle
%{\normalsize \small Email: \{shisheng.hu@uwaterloo.ca, jie.gao6@carleton.ca, k2qu@uwaterloo.ca, mushu1.li@ryerson.ca, c89zhou@uwaterloo.ca, sshen@uwaterloo.ca\}}}
%\author{Shisheng Hu, Yiyang Pei, Ying-Chang Liang, %\IEEEmembership{Fellow, IEEE}
%\thanks{S. Hu and Y.-C. Liang are with the Center for Intelligent Networking and Communications (CINC), University of Electronic Science and Technology of China (UESTC), Chengdu, 611731, China (e-mail: hss\_ss@163.com; liangyc@ieee.org).}
%\thanks{Y. Pei is with the Singapore Institute of Technology, Singapore, 138683,
%(e-mail: yiyang.pei@singaporetech.edu.sg).}}

%When a data drift occurs, a small deep neural network (DNN) on a computation-limited integrated sensing and communication (ISAC) device, that is used for sensing data processing, can suffer from accuracy degradation. In this case, a large amount of sensing data need to be offloaded to an edge computing server for processing by a large DNN, potentially resulting in transmission and server congestion.
\begin{abstract}
%Sensing data involved in integrated sensing and communication (ISAC) in 6G can be massive in volume and highly dynamic. %Accurately processing the generated sensing data with efficient utilization of distributed computing resources poses a significant challenge.
In this paper, we propose a digital twin (DT)-based user-centric approach for processing sensing data in \bl{an integrated sensing and communication (ISAC) system} with high accuracy and efficient resource utilization. The considered scenario involves an ISAC device with a lightweight deep neural network (DNN) and a mobile edge computing (MEC) server with a large DNN. \bl{After collecting sensing data}, the ISAC device either processes the data locally or uploads them to the server for higher-accuracy data processing. To cope with data drifts, the server updates the lightweight DNN when necessary, referred to as continual learning. Our objective is to minimize the long-term average computation cost of the MEC server by optimizing two decisions, i.e.,  sensing data offloading and sensing data selection for the DNN update. A DT of the ISAC device is constructed to predict the impact of potential decisions on the long-term computation cost of the server, based on which the decisions are made with closed-form formulas. Experiments on executing DNN-based human motion recognition tasks are conducted to demonstrate the outstanding performance of the proposed DT-based approach in computation cost minimization.

\end{abstract}
%To achieve this objective, a DT of the ISAC device is constructed to predict the accuracy of a retrained DNN and the time duration up to the next data drift. With the predictions, the benefit of the decisions on reducing the computation cost of the server can be evaluated.

\section{Introduction}
%1. What is ISAC and why it can assist ambient intelligence.

%2. What and why do we need edge continual learning

%3. What's the existing works about continual learning? what's the need of digital twin-based user-centric?

%4. contributions.

%Ambient intelligence aims to create smart environments that are sensitive to the presence of humans and responsive to their needs, thereby enhancing the experience in applications such as smart home and healthcare \cite{dunne2021survey}.

Integrated sensing and communication (ISAC), emerging as a promising element of the sixth generation (6G) networks, has the potential to offer ubiquitous sensing services by utilizing wireless infrastructures\cite{ISAC_survey,Jie_radar}. With artificial intelligence (AI) techniques, in particular deep neural networks (DNNs), \bl{sensing data collected in an ISAC system} can be processed to detect human presence and comprehend the surrounding environments. This enhances user experience in applications such as smart homes and smart factories \cite{edge_learning_sensing_survey}. To process the potentially massive sensing data, the distributed computing resources within a wireless network can be utilized. First, some of the sensing data can be offloaded to a mobile edge computing (MEC) server for processing by a large and highly accurate pre-trained DNN \cite{edge_learning_sensing_survey}. Second, a \textit{specialized} lightweight DNN \cite{yang2022dyco} can be deployed on an ISAC device (e.g., an user equipment with both communication and radar modules\cite{Networked_sensing}), which is trained to process sensing data of a relatively stationary distribution.

However, the sensing data collected by an ISAC device can follow a time-varying distribution. For example, when employing a time-division scheme in \bl{an ISAC system}, the duration allocated for sensing can be altered based on the communication demands of users, which leads to the change of the time dimension in the spectrogram of the sensing data. If the distribution of the currently collected sensing data diverges from that of the data used for training the lightweight DNN, a \textit{data drift} occurs, which can degrade the accuracy of the lightweight DNN on the ISAC device. In this case, the lightweight DNN on the ISAC device needs to be updated. The update can be achieved by either delivering a suitable lightweight DNN from the server if available \cite{Insitu_model_downloading} or having the server retrain the lightweight DNN based on the sensing data after the data drift\cite{drift_detection}. The latter method is referred to as edge continual learning.

%Sensing data generated in ISAC can have a large volume due to the prevalence of potential ISAC devices, e.g, wireless infrastructures and user equipment equipped with both communication and radar modules\cite{Networked_sensing}. In addition, the sensing data generated in ISAC can have a complex distribution. The complexity can result from the dynamic behaviors of users utilizing both communication and sensing services. First, users' communication throughput demands can fluctuate over time. This results in dynamic resource availability for sensing, such as the sensing duration in the time-division scheme in ISAC\cite{ISAC_survey}. %For example, when employing a time-division scheme in ISAC \cite{ISAC_survey}, the duration allocated for sensing can be altered.
%Second, users who utilize sensing services and act as sensing targets can display time-varying motions, distances, postures, and clothing. Both variations can affect the time or frequency dimensions of the spectrogram associated with the sensing data, adding to the complexity of its distribution.
%, which in turn leads to a change of the spectrogram of the sensing data.

Existing works on edge continual learning mainly focus on determining the selection (e.g., \cite{yang2022dyco}) and the scheduling (e.g., \cite{continous_edge_learning}) of the training samples for balancing the training cost and the enhanced accuracy of the retrained DNN. When the server is involved in both sensing data processing and lightweight DNN retraining, two challenges arise. First, since each sample selected by the server for retraining the lightweight DNN is offloaded from the ISAC device, the decisions on the sensing data offloading and training data selection are inherently coupled. Second, the accuracy of a retrained lightweight DNN is insufficient to characterize the gain of the DNN retraining since the retrained lightweight DNN will be replaced after the next data drift occurs. Moreover, it is difficult to evaluate the gain beforehand due to the dynamic and user-specific nature of both the accuracy (as a function of training data selection) and the data drift.

%we jointly optimize the decisions on sensing data offloading and training data selection, for minimizing the long-term average computation cost of an server for sensing data processing and DNN retraining. We use the digital twin (DT) paradigm \cite{shen2021holistic, glo_com_sshu, ConghaoDT} and construct a DT of the ISAC device. Simulation results show that based on the decision evaluation by DTs, the proposed approach can greatly reduce the computation cost of an server.

%In this paper, to tackle the first challenge, we jointly optimize the decisions on sensing data offloading and training data selection, for minimizing the overall edge computation cost for data processing and DNN retraining. To tackle the second challenge, we use the digital twin (DT) paradigm \cite{shen2021holistic, glo_com_sshu, ConghaoDT} and construct a DT of the ISAC device. The DT is used to predict the accuracy of the retrained small DNN and the next data drift, and accurately evaluating the sensing data offloading and training data selection decisions. Based on the evaluation, the original stochastic optimization problem is transformed, and an alternate convex search algorithm is proposed. In addition, the closed-form optimal solution of each decision given the other decision is derived for efficient decision-making.

In this paper, based on the digital twin (DT) paradigm \cite{shen2021holistic, glo_com_sshu, ConghaoDT}, we propose a novel user-centric approach in an ISAC system. Our objective is to minimize the long-term average computation cost of an MEC server for sensing data processing and DNN retraining by jointly optimizing the decisions on sensing data offloading and training data selection. The main contributions of this paper are summarized as follows:
\begin{itemize}
\item We develop a DT-based approach to user-centric edge continual learning, which can capture the user's unique time-varying communication demands and predict the gain of edge continual learning for the corresponding ISAC device. %to predict the accuracy of the retrained lightweight DNN and the occurrence of the next data drift. Then, the amount of sensing data offloaded by the ISAC device until the next data drift and the corresponding computation cost for processing the data are predicted.
\item We design an efficient algorithm for sensing data offloading and training data selection in edge continual learning based on the prediction results from the DT. Experiments on executing human motion recognition tasks are conducted to show that with the designed algorithm, the long-term average computation cost of the server can be largely reduced.
\end{itemize}

\section{System Model and Problem Formulation}\label{sec: system model}
\begin{figure}
    \centering
    \includegraphics[width=7.8cm]{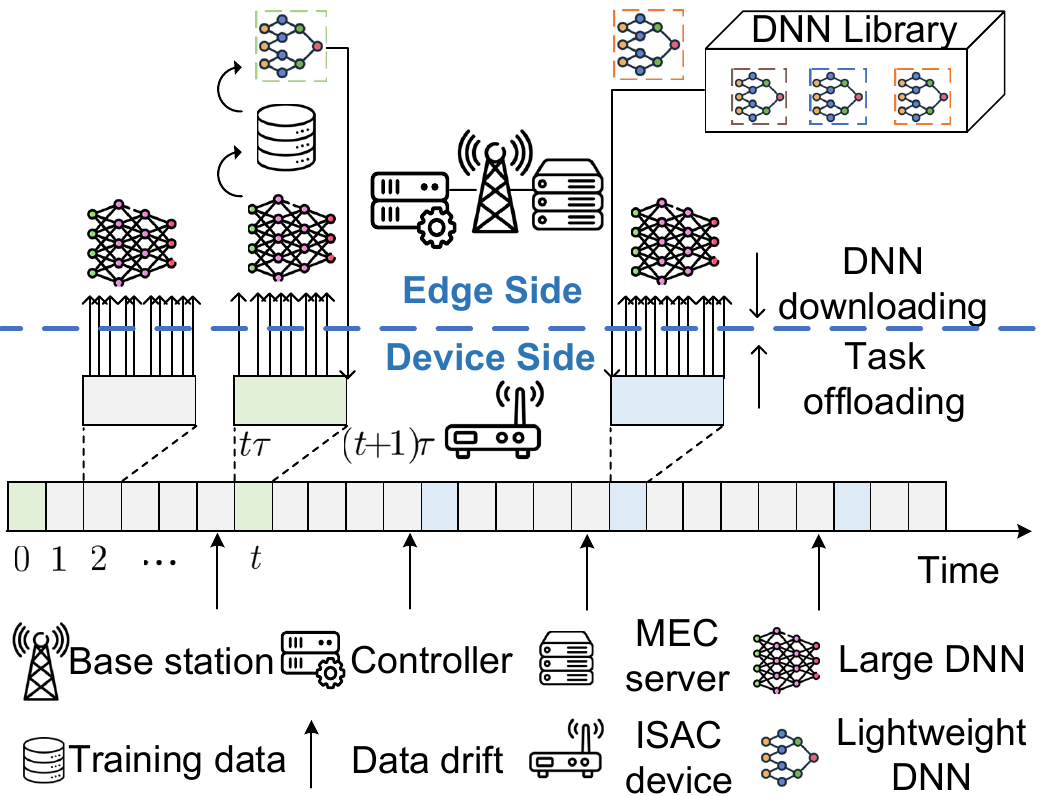}
    \caption{System model.}
    \label{system_model}
\end{figure}
\begin{figure}
    \centering
    \includegraphics[width=7.8cm]{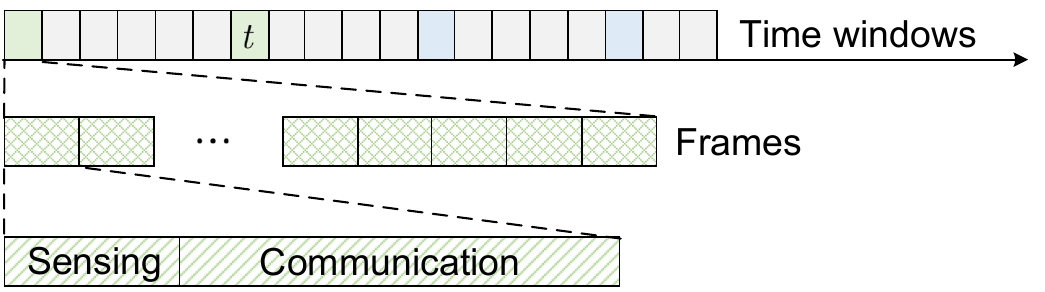}
    \caption{Illustration of time windows and frames.}
    \label{time_structure}
    \vspace{-0.6cm}
\end{figure}
\subsection{Network Model}
As shown in Fig. \ref{system_model}, we consider a scenario where an ISAC device, \bl{e.g., a user equipment with both communication and sensing modules}, is deployed in an area for sensing and communication. An MEC server is deployed co-located with a base station to assist the device in processing the sensing data. As shown in Fig. \ref{time_structure}, time is divided into time windows, each with a length of $\tau$ seconds, and the index of a time window is denoted by $t \in\{1,2,3,...\}$. Each time window consists of $N$ frames of the same length. Similar to \cite{ambient_sensing}, the time-division scheme is adopted by the ISAC device. Specifically, in each frame, the device first transmits a sensing signal and receives the echo signal from the target. In the remaining time of the frame, the device transmits communication signals. %The sensing data sample is then processed by a DNN to recognize the behaviour pattern of the sensing target e.g., motion and anomaly.

%The data distribution when the target is in different areas is different. For example, the SNR varies for echo signals reflected from different ranges.}. At the beginning of each time window, the device uses an algorithm (e.g., confidence level-based algorithm) to detect whether the data drift occurs. Define $I_t$ as the indicator of data drift detection, where $I_t$ equals to $1$ if data drift is detected in the $t$-th time window, and zero otherwise.

 %Then, a low-cost preprocessing method (e.g., threshold-based method) is applied to determine whether there is a target in the sensing range of the ISAC device. If there is a target detected, a sensing task is generated to process the sensing data sample for analyzing the behavior pattern of the target (e.g., motion detection and fall detection). The target arrival (and the sensing task generation) in the each time window is assumed to follow a Poisson process with a stationary arrival rate $\lambda$ in terms of tasks per second.

\subsection{Sensing Task Generation and Processing}
In each frame, \bl{the ISAC device preprocesses the received echo signal by extracting features from the echo signal}, such as a time-Doppler spectrogram \cite{ambient_sensing}, to create a sensing data sample in this frame. A sensing task is then generated by the ISAC device to process the sensing data sample by a DNN (e.g., human motion recognition and fall detection). Each sensing task can be either \bl{\textit{i)} offloaded to the MEC server to be processed by a large pre-trained DNN}, or \textit{ii)} locally processed by a lightweight DNN. We consider a probabilistic task offloading policy. Specifically, in the $t$-th time window, a sensing task generated in each frame is offloaded to the server with an \textit{offloading probability} of $\rho_t\in[0,1]$ and is locally processed with a probability of $1-\rho_t$. The average accuracy of the sensing tasks in the $t$-th time window, denoted by $A_t$, is calculated as:
\begin{equation}\label{ava_acc}
A_t = (1-\rho_t) A^{\rm D}_t+ \rho_t A^{\rm E},
\end{equation}
\bl{where $A^{\rm D}_t\in [0,1]$ represents the average accuracy of the on-device lightweight DNN} over sensing data samples in the $t$-th time window, and $A^{\rm E}$ represents the average accuracy of the large DNN on the server. The large DNN with massive parameters is offline trained by a large dataset of the labeled sensing data samples, and we assume that the large pre-trained DNN on the server is perfectly accurate, i.e., $A^{\rm E}= 1$. % \le A^{E} \le 1$.

Let $O_t^{\rm I}$ represent the expected computation cost of the server for processing the sensing tasks offloaded from the device in the $t$-th time window. It can be calculated as $O_t^{\rm I} = N\rho_tc^{\rm I}$, where $c^{\rm I}$ represents the average computation cost in terms of the number of floating-point operations (FLOPs) to process each sensing task by the large DNN at the server.

\subsection{Updating the Lightweight DNN on the ISAC Device}
\begin{figure}
    \centering
    \includegraphics[width=8cm]{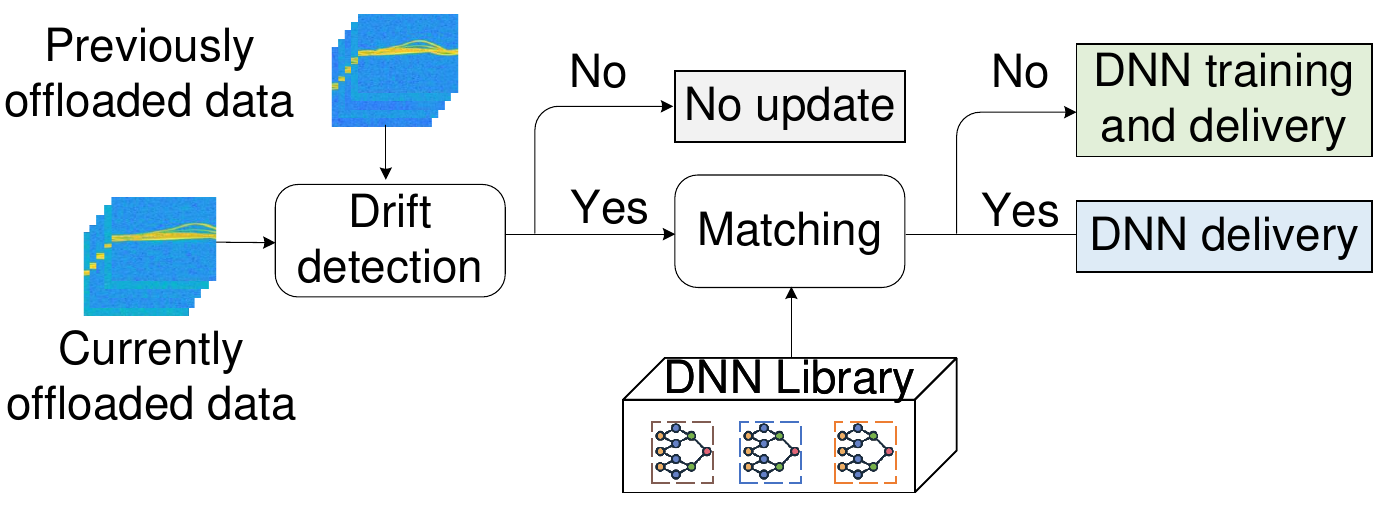}
    \caption{The procedure to determine whether and how to update the lightweight DNN at the ISAC device.}
    \label{DNN Updating}
    \vspace{-0.5cm}
\end{figure}

\bl{For an ISAC device, a data drift can occur due to the time-varying resource availability for sensing. For example, in smart home applications, the average available sensing time can change depending on whether the user is conducting activities with high communication demands such as video streaming.} The user's communication activities affect the device's sensing duration, which in turn may alter the distribution of the time-Doppler spectrogram of the sensing data. \bl{Without loss of generality, we assume that a data drift can occur at most once in any given time window.} If a data drift occurs, the distribution of the sensing data changes and then remains the same until the next data drift. %\footnote{The sudden data drift can also occur due to the movement of the sensing target. For example, a sensing target, such as an individual, moves within a specific area of the home for a duration before shifting to another area.}.
As shown in Fig. \ref{DNN Updating}, at the beginning of each time window, a drift detection algorithm (e.g., \cite{drift_detection}) is applied to determine whether a data drift occurs in the previous time window. If a data drift is detected, denoted by $I_t=1$, the lightweight DNN at the ISAC device needs to be updated in either of the following two cases. Otherwise, we have $I_t=0$.

\textit{1) When a matched DNN is absent:} The server hosts a DNN library, containing several pre-trained lightweight DNNs. Each lightweight DNN is trained using sensing data from a distinct data distribution. If no lightweight DNN in the DNN library is a match for the data distribution after a data drift, denoted by $M_t=0$ (represented by the green blocks in Fig. \ref{system_model} and Fig. \ref{DNN Updating}), the server will retrain the lightweight DNN for the ISAC device. In the first $\tau-\mu$ seconds of the $t$-th time window, after a sensing task is offloaded to and processed by the server using a large DNN, \bl{the sensing data and sensing result are selected to construct one training sample for retraining the lightweight DNN with a \textit{training probability} of $\gamma_t\in(0,1]$.} As a result, the expected number of samples used to retrain the lightweight DNN in the $t$-th time window is $\eta N\rho_t\gamma_t$, where $\eta = (\tau-\mu)/\tau$. In the final $\mu$ seconds of this time window, the server will retrain the lightweight DNN and deliver the retrained DNN to the ISAC device.\looseness=-1

\textit{2) When a matched DNN exists:} \bl{If a lightweight DNN in the DNN library is a match for the data distribution after the data drift}, which we denote by $M_t=1$ (represented by the blue blocks in Fig. \ref{system_model} and Fig. \ref{DNN Updating}), the server will deliver the DNN (with its accuracy known) to the ISAC device \bl{by the beginning of the next time window} \cite{Insitu_model_downloading}. In this case, we have $\gamma_t=0$. \looseness=-1% i.e., $A^{D}_t = A^D_0$ if $I_t=1$ and $M_t=0$.

An updated DNN will be used for the on-device processing of sensing tasks until the next data drift is detected. The expected computation cost of retraining the lightweight DNN in the $t$-th time window is denoted by $O_t^{\rm R}$ and can be calculated as $O_t^{\rm R} = \eta N\rho_t\gamma_tc^{\rm R}$, where $c^{\rm R}$ represents the average computation cost, in terms of the number of FLOPs, for using each training sample to retrain the lightweight DNN.

\subsection{Problem Formulation}
The expected accuracy of the sensing tasks in each time window should be higher than a threshold:
\begin{equation}\label{acc_constraint}
A_t \ge A^{\rm min}, ~\forall t\in\{1,2,3,...\},
\end{equation}
where $A^{\rm min}\in [A^{\rm D}_t,1]$. By combining (\ref{ava_acc}) and (\ref{acc_constraint}), we can obtain the minimum task offloading probability in the $t$-th time window, denoted by $\rho_t^{\rm min}$, that satisfies the average accuracy requirement:\looseness=-1
\begin{equation}\label{acc_equal_constraint}
\rho_t^{\rm min} = 1- \frac{1-A^{\rm min}}{1-A^{\rm D}_t}.
\end{equation}

In the beginning of each time window, a network controller determines the task offloading probability, $\rho_t$, and the training probability, $\gamma_t$, for the ISAC device in this time window. To support more ISAC devices, our objective is to minimize the average edge computation cost for the ISAC device over all time windows. The problem is formulated as:
\begin{subequations}\label{original_prob}
\begin{align}
{\rm(P1)}\bm{\min}_{\{\rho_t,\gamma_t, ~t=1,2,..., T\}} ~~~&\lim_{T \to \infty}\frac{1}{T}\sum_{t=1}^{T} \left(N\rho_tc^{\rm I}+ \eta N\rho_t\gamma_{t}c^{\rm R}\right),\label{obeject}\\
{\bm{\mathrm{s.t.~~~}}} &\rho_t^{\rm min} \le \rho_t \le 1,\label{P1constraint1}\\
&0\le \gamma_t \le \max\{(1-M_{t})I_{t},1\},\label{P1constraint2}
\end{align}
\end{subequations}
where (\ref{P1constraint1}) ensures that the minimum task offloading probability in (\ref{acc_equal_constraint}) should be attained to satisfy the accuracy requirement in (\ref{acc_constraint}). \bl{Constraint (\ref{P1constraint2}) ensures that the lightweight DNN is retrained by the server} in a time window only if \textit{i)} a data drift is detected in this time window, i.e., $I_t=1$, and \textit{ii)} there is no lightweight DNN in the DNN library that is a match for the current data distribution, i.e., $M_t=0$.

In the time windows when the server does not need to retrain the lightweight DNN for the ISAC device, the optimal task offloading probability equals the minimum task offloading probability calculated in (\ref{acc_equal_constraint}). To this end, in the following sections, unless specified otherwise, we consider the time windows in which $I_{t}=1$ and $M_{t}=0$.

\section{Digital Twin-Based User-Centric Approach for Edge Continual Learning}\label{sec: DT approach}

It is difficult to solve (P1) directly since two pieces of important information needed for this problem are unknown in advance: \bl{\textit{i)} whether a data drift will occur in a time window or not, for all the considered time windows}, and \textit{ii)} the accuracy of the lightweight DNN given a task offloading and training probability. To estimate the above information for accurately evaluating the decisions on the task offloading and training probabilities, we construct a DT of the ISAC device and propose a user-centric approach using the DT.\looseness=-1 %With the obtained information, we can evaluate the decisions in the long run.

%First, the accuracy prediction is important since the accuracy of the on-device DNN determines the minimum task offloading probability and thus affects the computation cost at the server in the future time windows for task processing. Second, the time instant of the occurrence of the next data drift would determine how many future time windows the on-device DNN which is retrained in a time window can be used.

%the relation of the accuracy and the number of used training samples in any time window, i.e., $(a_t,b_t)_{t=1}^{T}$. Note that decisions made in the beginning of the $t$-th time window will directly affect the computation cost in the $t$-th time window and minimum task offloading probability in the ($t+1$)-th time window. To address this issue, we proposed a DT-based solution.

\subsection{DT-Based Prediction of Data Drift Occurrence}
The moment when the next data drift occurs is important for solving (P1) as it determines the number of time windows for which retraining the lightweight DNN can be beneficial. Let $T_t$ represent the number of time windows up to the next data drift, and we have:
\begin{equation}
I_k=\left\{
\begin{aligned}
&0, ~{\rm if}~ t+1\le k\le t+T_t-1 ~{\rm and}~ T_t>1,
\\& 1, ~{\rm if}~ k=t+T_t.
\end{aligned}
\right.
\end{equation}

%The previous $W$ time intervals between two consecutive detected data drifts is used as the DT model of its expectation value.
Suppose that at the beginning of the $t$-th time window, $\begin{matrix}\sum_{k=1}^{t}I_k\end{matrix}$ occurrences of data drift have been observed. Accordingly,  $\begin{matrix}\sum_{k=1}^{t}I_k\end{matrix}-1$ time intervals between two consecutive data drifts can be obtained as a sequence $\mathcal{D}_t =\{D_1, D_2,...\}$, where $D_i$ represents the number of time windows between the $i$-th data drift and the ($i+1$)-th data drift, and $\vert \mathcal{D}_t\vert =\begin{matrix}\sum_{k=1}^{t}I_k\end{matrix}-1$. The average of the latest $W$ values in $\mathcal{D}_t$ is used as the DT model of the expected value of $T_t$, denoted by $\overline{T}_t$. At the beginning of the $t$-th time window, $\overline{T}_t$ is updated as: \begin{equation}
\overline{T}_t = \frac{\sum_{i=\max\{1, \vert \mathcal{D}_t\vert -W+1\}}^{\vert \mathcal{D}_t\vert}{D_{i}}}{\max\{\vert \mathcal{D}_t\vert , W\}}.
\end{equation}

\subsection{DT-Based Prediction of Lightweight DNN Accuracy}
\begin{figure}
    \centering
    \includegraphics[width=4.8cm]{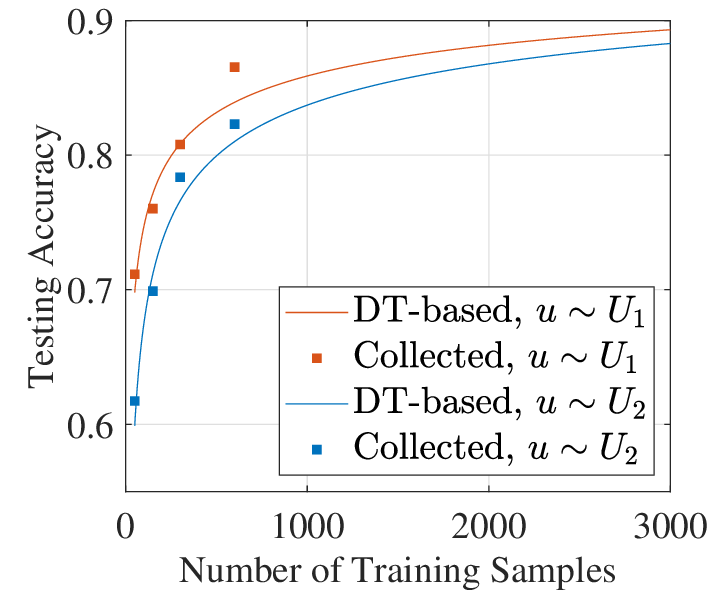}
    \caption{The testing accuracy of human motion recognition by a lightweight DNN versus the number of training samples used, where $u$ represents the sensing duration in a frame. See Section \ref{sec: simulation} for more details on the settings.}
    \label{acc_prediction_fig}
    \vspace{-0.5cm}
\end{figure}

The accuracy of the retrained lightweight DNN is also important for solving (P1) \bl{since it determines the minimum task offloading probability in (\ref{acc_equal_constraint}) for the next $T_t$ windows.} The accuracy is generally non-decreasing with the number of training samples used to train the DNN, \bl{which is determined by $\rho_t$ and $\gamma_t$ in our considered problem.} Such a relation can be time-varying based on the characteristic of the sensing data. For example, in Fig. \ref{acc_prediction_fig}, it is shown that with the sensing duration following two different uniform distributions, the corresponding accuracy of a lightweight DNN retrained by using the same number of training samples can be different. %To characterize such a time-varying relation for wisely selecting the task offloading and training probability, the DT of the ISAC device is constructed at the server.

\textit{1) DT Model:} Similar to \cite{Learning-centric_expoential}, an exponential function is used to predict the accuracy of the retrained lightweight DNN:
\begin{equation}
A_{k}^{\rm D} (x) = 1 - a_t x^{-b_t}, t+1\le k \le t+T_t,
\end{equation}
where $A_{k}^{\rm D}(x)$ represents the accuracy of the lightweight DNN that is trained by $x$ samples in the $t$-th time window, and $a_t$ and $b_t$ are positive learnable parameters.

\textit{2) DT Data Collection and DT Model Update:} From the training process in the $t$-th time window, $K_t$ samples are collected:
\begin{equation}
\mathcal{N}_{t} = \left\{(x_{1},A_{{t},1}),(x_{2},A_{{t},2}),...,(x_{K_t},A_{{t},{K_t}})\right\},
\end{equation}
where $A_{{t},i}$ represents the accuracy of the retrained lightweight DNN with $x_i$ training samples in the $t$-th time window (e.g., $K_t = 4, x_1=50, x_2=150, x_3 = 300$, and $x_4=600$ in Fig. \ref{acc_prediction_fig}).\looseness=-1

The parameters in the DT model are updated at the beginning of the $(t+1)$-th time window, by minimizing the mean square error of the accuracy prediction on the samples $\mathcal{N}_t$:\looseness=-1
\begin{equation}\label{DT_update}
(a_{t+1}, b_{t+1}) =  \arg \min_{a_{t+1},b_{t+1}} \sum_{i=1}^{\vert \mathcal{N}_t\vert }{\left(1\!-\! a_{t+1}x_i^{-b_{t+1}}\!-\!A_{t,i}\right)^2}.
\end{equation}

\textit{3) DT-Based DNN Accuracy Prediction:} Leveraging the DT, the accuracy of the retrained lightweight DNN in the $k$-th time window, i.e., $A_{k}^{\rm D} (\rho_t,\gamma_t)$, given the task offloading and training decisions in the $t$-th time window, i.e., $\rho_t$ and $\gamma_t$, can be predicted as:
\begin{equation}\label{acc_prediction}
A_{k}^{\rm D} = 1 - a_t(\eta N\rho_t\gamma_t)^{-b_t}, t+1\le k \le t+T_t.
\end{equation}
Then, the minimum task offloading probability in the $k$-th time window, where $t+1\le k \le t+T_t$, given by (\ref{acc_equal_constraint}), can be calculated as:
\begin{equation}\label{est_min_off}
\begin{aligned}
\overline{\rho}_{k}^{\rm min}(\rho_t,\gamma_t)&=1- \frac{1-A^{\rm min}}{a_t(\eta N\rho_t\gamma_t)^{-b_t}}\\
&\triangleq \overline{\rho}_{0,t}^{\rm min}(\rho_t,\gamma_t), t+1\le k \le t+T_t.
\end{aligned}
\end{equation}

\subsection{DT-Based User-Centric Decision Evaluation}
In the beginning of the $t$-th time window, with the prediction of the next data drift, we can estimate the expected number of time windows when the retrained lightweight DNN can be used, i.e., $\overline{T}_t$. With the DT-based prediction of the lightweight DNN's accuracy, we can estimate the minimum task offloading probability when the retrained lightweight DNN is used, i.e., $\overline{\rho}_{0,t}^{\rm min}(\rho_t,\gamma_t)$ in (\ref{est_min_off}). As a result, we can estimate the expected computation cost of task processing for the edge server until the next data drift as $ N\rho_tc^{\rm I}+\overline{T}_t N \overline{\rho}^{\rm min}_{0,t}(\rho_t, \gamma_t)c^{\rm I}$. Since retraining the lightweight DNN at the device can reduce the volume of sensing data to offload and in turn the associated edge computation cost, the decrease of this cost is the gain of DNN retraining. Adding this expected cost of task processing to the cost of DNN retraining, i.e., $N\rho_t\gamma_{t}c^{\rm R}$, we define the long-term computation cost resulting from decisions $\rho_t$ and $\gamma_t$, denoted by $O_t(\rho_t, \gamma_t)$ and calculated as:%\footnote{\bbl{As DNN retraining can reduce the edge computation cost for sensing data processing while incurring additional edge computation cost, the overall cost can be used to evaluate the DNN retraining.}}:
\begin{equation}
O_t(\rho_t, \gamma_t) = N\rho_t\gamma_{t}c^{\rm R}+ N\rho_tc^{\rm I}+ \overline{T}_t N\overline{\rho}^{\rm min}_{0,t}(\rho_t, \gamma_t)c^{\rm I}.
\end{equation}
%As DNN retraining can reduce the edge computation cost for sensing data processing while incurring additional edge computation cost, the cost can be used to evaluate the DNN retraining.
\section{Problem Transformation and Solution}\label{sec: problem trans}
%where $\overline{T}_t$ represents the expected number of time windows during which the data distribution remains unchanged.

With the DT-based prediction results, we transform the long-term optimization problem into consecutive one-shot optimization problems. Specifically, in the beginning of a time window when the lightweight DNN needs to be retrained, the optimal task offloading and training probabilities in this time window should minimize the long-term computation cost:
\begin{subequations}\label{transformed_prob}
\begin{align}
{\rm(P2)} \bm{\min}_{\rho_t,\gamma_t}~~~ &O_t(\rho_t,\gamma_t)\label{object}\\
{\bm{\mathrm{s.t.~~~}}}&\overline{\rho}_t^{\rm min}\le \rho_t \le 1,\label{P2constraint1}\\
&0\le \gamma_t \le 1.\label{P2constraint2}
\end{align}
\end{subequations}

%The second derivative of $O_t$ the with respect to the task offloading probability

\subsection{Properties of the Transformed Problem}
\textit{\textbf{Proposition 1:}} If $0< b_t< 1$, $O_t(\rho_t,\gamma_t)$ is a bi-convex function.
\begin{proof}
With $\gamma_t$ fixed, the second partial derivative of $O_t$ with respective to $\rho_t$ is:
\begin{equation}\label{second_deriative_rho_t}
\begin{aligned}
\frac{\partial^2 O_t}{\partial (\rho_t)^2}= \frac{1-b_t}{a_t}\underbrace{\overline{T}_t\gamma_t^2b_t(1-A^{\rm min})c^{\rm I}\eta^2N^3(\eta N\rho_t\gamma_t)^{-2+b_t}}_{\ge 0}.
\end{aligned}
\end{equation}

With $\rho_t$ fixed, the second derivative of $O_t$ with respective to $\gamma_t$ is:
\begin{equation}\label{second_deriative_gamma_t}
\begin{aligned}
\frac{\partial^2 O_t}{\partial (\gamma_t)^2} = \frac{1-b_t}{a_t}\underbrace{\overline{T}_t\rho_t^2b_t(1-A^{\rm min})c^{\rm I}\eta^2N^3(\eta N\rho_t\gamma_t)^{-2+b_t}}_{\ge 0},
\end{aligned}
\end{equation}
where the highlighted parts are positive since $A^{\min}\le 1$.

As a result, if $0< b_t< 1$, the two partial second derivatives are both positive.
\end{proof}

\textit{\textbf{Proposition 2:}} If $b_t\ge 1$, $O_t(\rho_t,\gamma_t)$ is a bi-concave function.
\begin{proof}
From (\ref{second_deriative_rho_t}) and (\ref{second_deriative_gamma_t}), it can be observed that when $b_t\ge 1$, the second partial derivatives are both non-positive.
\end{proof}

\subsection{Algorithm}

Based on the propositions, Algorithm \ref{ACS_algo} is proposed to efficiently solve (P2). Specifically, when $0< b_t<1$, since the objective function of is a bi-convex function and the constraints are convex, (P2) is a bi-convex optimization problem. In this case, an alternate convex search method \cite{gorski2007biconvex} is used in Algorithm \ref{ACS_algo}. When $b_t\ge1$, the objective function of (P2) is a bi-concave function. In this case, the minimizer of the function can be obtained at the Cartesian product of the vertex sets of (\ref{P2constraint1}) and (\ref{P2constraint2}) \cite[Theorem 3.10]{gorski2007biconvex}.

\subsection{Closed-Form Solutions}
If $0 < b_t< 1$, the closed-form solutions in each iteration of Algorithm \ref{ACS_algo} can be obtained. \bl{The partial derivatives of $O_t(\rho_t, \gamma_t)$ with respective to $\rho_t$ and $\gamma_t$ can be calculated by:}
\begin{equation}\label{first_deriative_rho_t}
\frac{\partial O_t}{\partial \rho_t} = Nc^{\rm I}\!+\!N\gamma_tc^{\rm R}\!-\!T_t\frac{(1\!-\!A^{\rm min})b_t c^{\rm I}\eta N^2 \gamma_t(\eta N\rho_t\gamma_t)^{-1\!+\!b_t}}{a_t},
\end{equation}
\begin{equation}\label{first_deriative_rho_t}
\frac{\partial O_t}{\partial \gamma_t} = Nc^{\rm R}\rho_t-T_t\frac{(1-A^{\rm min})b_t c^{\rm I}\eta N^2 \rho_t(\eta N\rho_t\gamma_t)^{-1+b_t}}{a_t}.
\end{equation}
Given the offloading probability $\rho_t$, the training probability $\gamma_t^{o}$ at which $\frac{\partial O_t}{\partial \gamma_t}|_{\gamma_t=\gamma_t^{o}}=0$ can be calculated as:
\begin{equation}\label{first_deriative_gamma_t_0}
\gamma_t^o= N^{\frac{b_t}{1-b_t}}\rho_t^{-1}\bigg(\frac{a_tc^{\rm R}}{c^{\rm I}\overline{T}_t\eta(1-A^{\rm min})b_t}\bigg)^{\frac{1}{-1+b_t}}.
\end{equation}
Given the training probability $\gamma_t$, the offloading probability $\rho_t^{o}$ at which $\frac{\partial O_t}{\partial \rho_t}|_{\rho_t=\rho_t^{0}}=0$ can be calculated as:
\begin{equation}\label{first_deriative_rho_t_0}
\rho_t^o = N^{\frac{b_t}{1-b_t}}(\eta\gamma_t)^{-1}\bigg( \frac{a_t(1+\gamma_tc^{\rm R})}{c^{\rm I}\overline{T}_t\eta(1-A^{\rm min})b_t\gamma_t}\bigg)^{\frac{1}{-1+b_t}}.
\end{equation}

\textit{\textbf{Proposition 3:}} The optimal training probability that minimizes the long-term computation cost, denoted by $\gamma_t^{\rm opt}$, when the offloading probability is $\rho_t$, can be calculated as:
\begin{equation}\label{closed-form_sol_gamma_t}
\gamma_t^{\rm opt}=\left\{
\begin{aligned}
&\gamma_t^{o}, ~{\rm if}~ 0\le \gamma_t^{o}\le 1,
\\& 1, ~{\rm otherwise}.
\end{aligned}
\right.
\end{equation}

The optimal task offloading probability that minimizes the long-term computation cost, denoted by $\rho_t^{\rm opt}$, when the training probability is $\gamma_t$, can be calculated as:
\begin{equation}\label{closed-form_sol_rho_t}
\rho_t^{\rm opt}=\left\{
\begin{aligned}
&\rho_t^{\rm min}, ~{\rm if}~ \rho_t^{o}\le \rho_t^{\rm min},
\\&\rho_t^{o}, ~{\rm if}~ \rho_t^{\rm min}\le\rho_t^{o}\le 1,
\\& 1, ~{\rm otherwise}.
\end{aligned}
\right.
\end{equation}

\begin{proof}
Omitted here due to space limitation.
\end{proof}

\textbf{\textit{Remark 1:}} Since $\frac{1}{-1+b_t}$ in (\ref{first_deriative_gamma_t_0}) is negative, the optimal training probability in (\ref{closed-form_sol_gamma_t}) increases with the expected number of time windows up to the next data drift, i.e., $\overline{T}_t$. This is expected since in this case a retrained lightweight DNN can be used for more time windows on average, and it is thus advantageous to increase the accuracy of the lightweight DNN by constructing and using more training samples.

\addtolength{\topmargin}{0.05in}
\renewcommand{\algorithmicrequire}{\textbf{Input:}}
\renewcommand{\algorithmicensure}{\textbf{Output:}}
\newcommand{\algorithmicbreak}{\textbf{break}}
\newcommand{\BREAK}{\STATE \algorithmicbreak}
\vspace*{0.05in}
\begin{algorithm}
\caption{Joint Optimization of Sensing Task Offloading and Selection for Edge Continual Learning} \label{ACS_algo}
{\small{
{
\begin{algorithmic}[1]   
\setlength{\baselineskip}{0.88\baselineskip}
%\STATE $\tau_s^{(0)},\tau_m^{(0)}$
%
\IF{$0< b_t< 1$}
\STATE $i \gets 0$
\STATE Initialize $\rho_t^{(0)}$ by randomly choosing from $[\rho^{\rm min}_t,1]$.
%\STATE Initialize $\gamma_t^{(0)}$ by randomly choosing from $[0,1]$.
%\REQUIRE $n,v$\
\REPEAT%{\textbf{True}}
    \STATE $i \gets (i+1)$
    \STATE Given $\rho_t^{i-1}$, find optimal $\gamma_t^{i} \in [0,1]$ that minimizes $O_t(\rho_t^{i-1},\gamma_t^{i})$ based on (\ref{first_deriative_gamma_t_0}) and (\ref{closed-form_sol_gamma_t}).
    \STATE Given $\gamma_t^{i}$, find optimal $\rho_t^{i}\in [\rho^{\rm min}_t,1]$ that minimizes $O_t(\rho_t^{i},\gamma_t^{i})$ based on (\ref{first_deriative_rho_t_0}) and (\ref{closed-form_sol_rho_t}).
\UNTIL $|\rho_t^{(i)}-\rho_t^{(i-1)}|$ and $|\gamma_t^{(i)}-\gamma_t^{(i-1)}|<10^{-2} $
\STATE$\rho_t^* \gets\rho_t^{(i)}$,$\gamma_t^* \gets\gamma_t^{(i)}$
\ENDIF
\IF{$b_t\ge 1$}
\STATE $(\rho_t^*, \gamma_t^*) = \arg \min_{\rho_t\in\{\rho_t^{\rm min},1\},\gamma_t\in\{0,1\}} O_t(\rho_t, \gamma_t)$
\ENDIF
\ENSURE $\rho_t^{*},\gamma_t^{*}$
\end{algorithmic}}}}
\end{algorithm}

\section{Simulation Results}\label{sec: simulation}
\vspace{-0.15cm}
In this section, we evaluate the performance of the proposed DT-based edge continual learning for a radar sensing-based human motion recognition task in \cite{ambient_sensing}, which classifies human motion into \textit{standing}, \textit{child pacing}, \textit{child walking}, \textit{adult pacing}, and \textit{adult walking}. The settings of the radar sensing parameters, including the bandwidth, carrier frequency, sweep time of a frequency-modulated continuous wave (FMCW) sensing signal, and the height and speed of the sensing target are set following \cite{ambient_sensing}. %The ISAC device is located at the coordinate \((0\text{m}, 0\text{m})\). The initial position of the sensing target is within a central rectangular area of size \(2\text{m} \times 3\text{m}\), and the distance from its central point to the ISAC device is \(3.5\text{m}\). The transmit power of the sensing signal is $20$ dBm and the power of additive ground clutter and receiver noise is $-104$ dBm \cite{Noise_setting}.
The signal reflected from a sensing target is simulated using the Phased Array System Toolbox of MATLAB. The received signal undergoes the following preprocessing steps, \textit{i)} sampling \bl{(at the frequency given by \cite{ambient_sensing})}, \textit{ii)} mix with a reference signal, \textit{iii)} short-time Fourier transform (STFT), and \textit{iv)} normalization and conversion into a $256\times256$ spectrogram. Each lightweight DNN consists of one convolutional layer and two fully-connected layers, and the computation cost for training it using each sample, i.e., $c^{\rm R}$, is $24.9$ GFLOPs \cite{epoch2022estimatingtrainingcompute}. ResNet-18 is selected as the large DNN, and the computation cost of using it for processing each task, i.e., $c^{\rm I}$, is $4.5$ GFLOPs.

The occurrence of a data drift in the $t$-th time window follows a Bernoulli distribution with the parameter $P^{\rm dr}_t$. We consider in total $T=1\times10^5$ time windows with $P^{\rm dr}_t=P^{\rm dr}_0, \forall t=1,2,3,...,T/2$, and $P^{\rm dr}_t=P^{\rm dr}_0/4,  \forall t=T/2+1,..., T$. In each time window, there are $N=1500$ frames and $\eta=0.8$. In addition, $M_t$ follows a Bernoulli distribution with the parameter $P^{\rm md}=0.75$. Moreover, the relation between the accuracy of a retrained lightweight DNN and the number of the training samples varies over time following a two-state Markov process. The transition matrix of the Markov process is $[\sigma, 1-\sigma;1-\sigma,\sigma]$, where $\sigma=0.3$ represents the probability that the relation remains unchanged when the lightweight DNN needs to be retrained. A state of the relation is described by the parameters $(a_t,b_t)$ \bl{in the DT model for predicting the accuracy in (\ref{acc_prediction})}. As shown in Fig. \ref{acc_prediction_fig}, two sets of the parameters are obtained when the sensing time in each frame follows $U(0.6, 0.8)$ and $U(0.4, 0.8)$ seconds, and the values are $(0.81, 0.25)$ and $(1.30,0.30)$ based on (\ref{DT_update}).

We consider the following three benchmarks.
\begin{itemize}
\item \textbf{Lower Bound:} The occurrence of a data drift in all the time windows and the relation between the accuracy of a retrained lightweight DNN and \bl{the number of training samples are perfectly known a priori.}
\item \textbf{Without DT Update:} The parameters for DNN accuracy prediction and the expected time interval between two consecutive data drifts are first estimated and then assumed unchanged. %In addition, the expectation of the time interval between two data drifts are updated based on the first $W=100$ time windows and are assumed unchanged. %, i.e., $\overline{T}_t = T_W, t=0,1,2,...$.
\item \textbf{Without DT:} When the lightweight DNN needs to be retrained, the minimum task offloading probability is selected, i.e., $\rho_t=\rho_t^{\rm min}$, and all the offloaded sensing tasks are used for retraining, i.e., $\gamma_t=1$. %\item \textbf{Pure Data-driven:} A deep reinforcement learning (DRL) algorithm is applied to directly output the task offloading probability and training probability.
%\item \textbf{Data and DT Co-driven:} A deep reinforcement learning (DRL) algorithm is applied to directly output the task offloading probability and training probability.
\end{itemize}

\begin{figure}
    \centering
    \includegraphics[width=4.6cm]{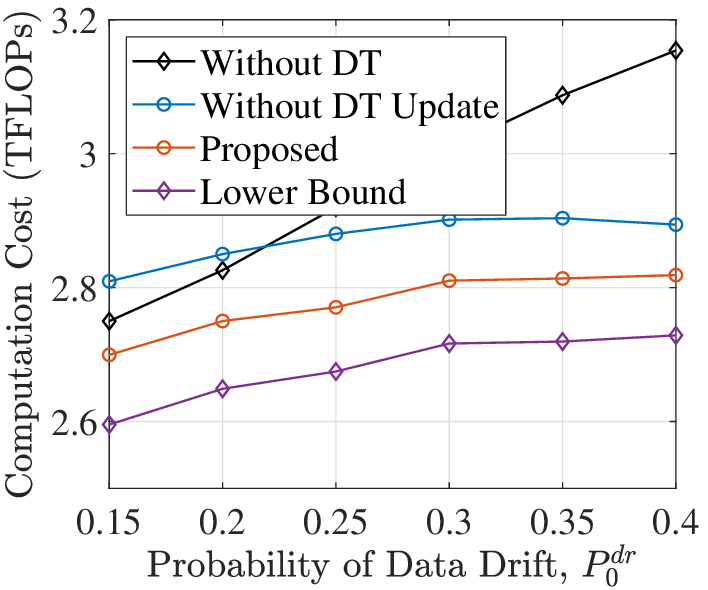}
    \caption{Average computation cost of the server versus the probability of data drift $P^{\rm dr}_0$.}
    \label{ava_comp_cost}
    \vspace{-0.65cm}
\end{figure}

\begin{figure}
    \centering
    \includegraphics[width=4.6cm]{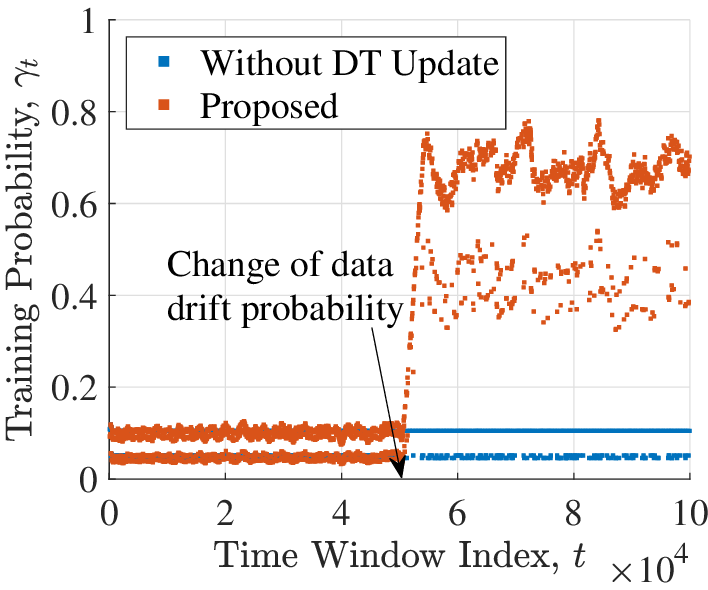}
    \caption{Training probability versus time window index.}
    \label{train_prob}
    \vspace{-0.65cm}
\end{figure}

In Fig. \ref{ava_comp_cost}, the average edge computation cost per time window versus the probability of the data drift, i.e., $P^{\rm dr}_0$, is shown. It can be observed that the proposed DT-based approach outperforms \bl{both the approach without DT and that without DT update}. In addition, the performance gain over the approach without DT increases with the probability of data drift occurrence. This is expected as the proposed approach uses the DT to estimate the time duration when a retrained lightweight DNN can be used in determining the training probability, while the approach without the DT uses \bl{all the offloaded sensing tasks and their results} to retrain the lightweight DNN. In Fig. \ref{train_prob}, given $P^{\rm dr}_0=0.3$, the training probability versus the index of the time window is shown. It can be observed that by the proposed approach, in the second half of the time windows, the training probability is increased. This is expected as the probability of a data drift is set to decrease after the ($T/2+1$)-th time window, and the expected time interval between two consecutive data drifts thus increases. Based on \textit{Remark 1}, the optimal training probability should increase in this case. By comparison, without updating the DT to capture such a change, the training probability is not appropriately adjusted.
\vspace{-0.2cm}

\section{Conclusion}\label{sec: conclusion}
\vspace{-0.2cm}
We have proposed a DT-based approach for user-centric edge continual learning in an ISAC system. A DT of an ISAC device has been constructed to capture the corresponding user's unique time-varying communication demands and their impact on the gain of edge continual learning. Then, sensing data offloading and training data selection have been optimized for the user, subject to the user's task processing accuracy requirement, while minimizing the computation cost for the MEC server. The proposed approach to user-centric edge continual learning is a step toward seamless integration of sensing, communication, and computation in 6G networks. For future work, we will investigate DT-based joint sensing, communication, and computing resource reservation for edge continual learning in a multi-device scenario.
%Thereby, accurately evaluating the decisions on sensing data offloading and training data selection in edge continual learning.
%\begin{figure}
%    \centering
%    \includegraphics[width=8.0cm]{tct_vs_f.eps}
%    \caption{Transaction confirmation time vs. the transaction fee in the transaction $f$.}
%    \label{UNversusf}
%\end{figure}
\bibliographystyle{IEEEtran}
\bibliography{Hss_Ref}
\vspace{-0.2cm}
\end{document}